# Organizing Family Support Services at ACM Conferences

Audrey Girouard, Carleton University
Jon E. Froehlich, University of Washington
CHI 2018 Family Chairs

Regan Mandryk, University of Saskatchewan
Mark Hancock, University of Waterloo
CHI 2018 General Chairs

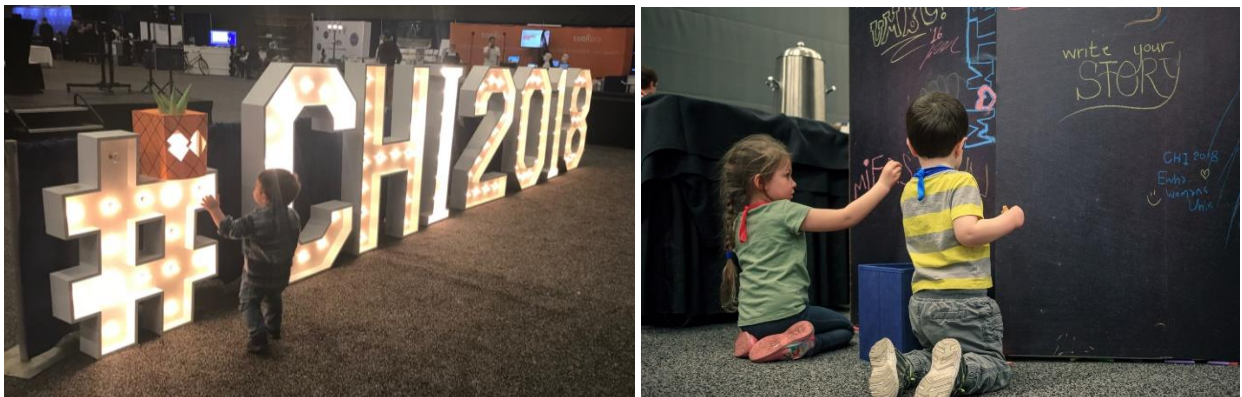

**Figure 1.** Two photos of young children at CHI2018 where professional childcare services were offered to support attendees with families.

## Introduction

Conferences matter. They offer a one-stop forum for academics and industry professionals to communicate recent findings, meet new people, foster collaborations, maintain connections, and nurture a sense of community. Conference attendees who are also parents, however, face a notable barrier: how to attend and gain the aforementioned career benefits while also balancing childcare responsibilities—a challenge that Calisi *et al.* [2] refer to as the "childcare-conference conundrum." While all primary caretakers of children are affected, women often experience greater disadvantages due to multiple factors (*e.g.,* biological, cultural) further influencing their participation in our communities and, ultimately, their careers. Indeed, recent work studying family formation on academic careers found that a "baby penalty" negatively affected women's but not men's career mobility, with even a larger negative impact for women of color [4].

To help address these issues, increasingly, academic and industry conferences are attempting to offer some sort of family-support services (*e.g.,* on-site nursing rooms and/or childcare) [1]. In this article, we describe planning, organizing, running, and assessing family-support services at ACM CHI 2018—a large multidisciplinary conference of researchers and practitioners in human-computer interaction and design. We write from our perspectives and experiences as the conference family co-chairs (Audrey, Jon) and general chairs (Regan, Mark) as well as academic parents.

Our goal with this article is threefold: first, to provide insights into the logistics, policies, and complexities of offering family services at a large (*N*=3,500) multidisciplinary conference; second, to describe key findings related to the needs and impact of family support services on our community; finally, to reflect on and offer guidelines and best practices for future ACM conferences and beyond.

This article is a complement to an abbreviated version published in the April 2020 issue of the Communications of the ACM. This longer version includes significantly more information on our approach to planning family services at CHI2018, decision making, and our pre-conference and post-conference assessments.

### History of Services Offered at CHI

Unlike many academic conferences, CHI has a rich, though inconsistent, history of offering family support services. In 1996, Allison Druin started the CHIkids program—a fun, interactive "CHI daycamp" based on Druin's research in children-focused participatory design [3]. In CHIkids, children helped brainstorm and mockup interaction designs, tried emerging prototypes, interviewed attendees, put out a daily newsletter, and produced a video for the closing plenary. To help run CHIkids, Druin contracted a professional childcare service called KiddieCorp and invited members of the CHI community to volunteer as "CHIkids Leaders" who were partially compensated with airfare, accommodations, and/or conference registration. To help cover childcare costs and to provide computer equipment to the children, Druin sought out and received external sponsorships from companies like Disney Interactive, Microsoft, and Brøderbund.

Because detailed records are not available for these early initiatives, it is not possible to comprehensively analyze demand, successful outcomes, painpoints, and cost; however, an archived CHIkids page from 1999[1] indicates 38 children enrollees (ages 6 months to 14 years old). An additional archived page from CHI2001[2] discusses two childcare options: one for $85/day intended for 6 months to 6 year olds and the other called "CHICamp" for $130/day for 7–14 year olds (which amounts to $121/day and $185/day in today's dollars, respectively). Services were offered from 8:30AM-6:00PM. Though CHIkids and its later incarnation CHICamp required a significant grassroot effort to organize and run, the program continued for nine years (under Druin's leadership from 1996–2001 and Sabrina Lao's from 2002–2005).

While some community-based initiatives occurred in the interim (*e.g.,* social media groups to coordinate babysitters), CHI did not offer on-site childcare again until 2016 when Jofish Kaye and Druin co-chaired the conference. This re-introduction of services was, in part, a reaction to certain previous flashpoints in which children were unable to accompany a parent into a conference social function (*e.g.,* the conference reception) but also seen by Kaye and Druin as an important initiative to increase the inclusiveness of the CHI conference. To organize and formalize the effort, Kaye and Druin created a childcare chair position within the CHI organizing committee and asked Louise Barkuus and Judd Antin to serve in the inaugural role. While a symbolic success in highlighting family inclusivity to the CHI community, enrollment did not meet expectations (approximately 15 children participated). Consequently, childcare services were not provided at CHI 2017 and no one was appointed to the role of family services chair. It is with this context that we began planning for CHI 2018.

## Our Approach to Planning Family Support Services

To plan for CHI 2018, we began with a background survey soliciting feedback about potential family support and childcare options at the conference. Concurrently, we inquired with conference chairs and colleagues who had experience providing childcare services at academic conferences.

---

[1] http://old.sigchi.org/chi99/chikids/chikids/index.html
[2] http://prior.sigchi.org/chi2001/facilities/index.html, http://prior.sigchi.org/chi2001/ap/additional/chicamp.html

## Pre-Conference Survey

To recruit a broad range of respondents, we advertised the survey on social media (*e.g.,* Facebook's CHI Meta, CHI Women and HCI Parents pages, the CHI Twitter account) and via mailing lists (*e.g.,* CHI-announcements). The survey was available for one month in October 2017, approximately 6 months before the conference. It contained a mixture of open- and closed-form questions. Not all questions were required.

In all, we had 95 respondents, including 66 faculty, 17 students, and 10 people from industry. Of these, 56 (58.9%) reported that they would be *more likely* to attend CHI 2018 if childcare services were offered and 66 (69.5%) indicated that they would be *likely* or *very likely* to use childcare services at CHI 2018. Note, however, that these numbers include 16 respondents who indicated that they themselves do not have children or do not require childcare. The analysis below is limited to those 79 respondents (83%) who stated needing childcare at CHI 2018.

We report data from questions inquiring about specific aspects of childcare and child-friendly options at the conference. At the end of the survey, we also had an open-form question soliciting general feedback/comments, to which 27 respondents provided a response. Most common was a small note to indicate explicit support for having family support services. For example, "*I definitely think CHI should be a community to support families,*" "[I] *may not need this personally but I highly value the availability of this to the community,*" and "*Great initiative. I hope you can figure something out.*" However, there were a few other interesting ideas/anecdotes, which we include when reporting specific question topics.

**Total children and demographics.** Our respondents estimated that they need childcare services for 108 children at CHI 2018, including: 23 babies (up to the age of one), 45 toddlers (between 1 and 4), 20 pre-school age children (between 4 and 6), and 20 school age kids (older than 6, with the oldest being 13).

Age is an important characteristic because: (i) there are typically governmental regulations controlling the ratio between providers and children, and more providers are needed for younger children; (ii) providing for nursing babies has more demanding facility requirements, including equipment to refrigerate and warm breast milk or formula. Also, on-site care would enable nursing mothers to more easily "drop-in" to nurse their child, which would decrease the chance of missing sessions or other conference related events.

At CHI 2016, the youngest age supported for on-site childcare was six months. While determining the appropriate age ranges to support is a complex decision, we received the following comment over email: "*I wanted to use the childcare service at CHI 2016 and wasn't able to because [my baby] was only three months old and childcare started at 6 months. So if you are able to offer childcare services, I think it would be important to especially provide it for children _under_ 6 months, because they are the ones who are most likely still nursing and thus need to travel with the parents.*" Two respondents indicated bringing kids 0–6 months, with five more being approximately at the 6-month mark. We also took into account that there might be some soon-to-be parents who might make a decision regarding bringing their future 0–6 month old child on a last-minute basis.

**Preferences for childcare services.** When asked about which childcare services would be of most interest (a select-all-that-apply closed-form question), 73 respondents (92.4%) selected *on-site childcare at the conference center (similar to CHI 2016 in San Jose)* followed by *on-site shared nannies at the conference center* (62%), and *independent babysitters/nannies* (38%). Figure 1 outlines the full list of preferences. As a write-in idea, one respondent suggested a "*room where parents can hang out with their small children, like a drop-in with some toys and a TV.*"

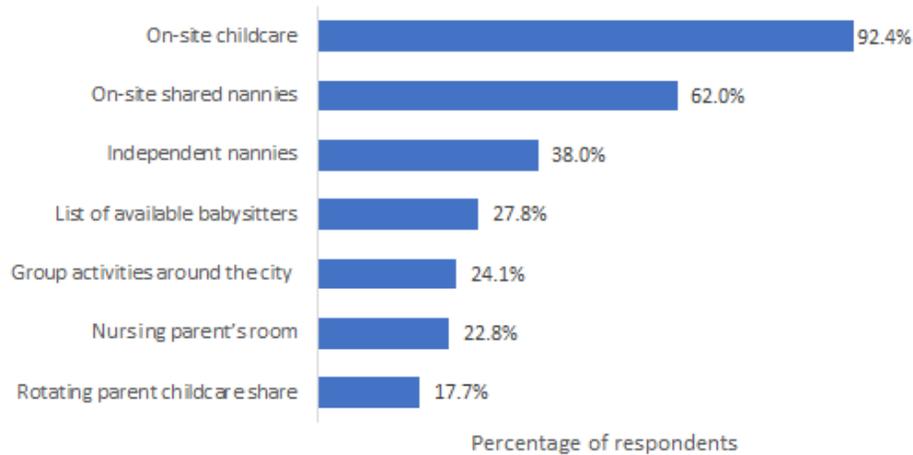

**Figure 1. Preferences for childcare services ($N$=95)**

With regards to the nursing room, two respondents shared some explicit thoughts:

*"I've nursed in some pretty awkward places at CHI, including an SV storage closet, dead-end hallways, and empty conference rooms. If a dedicated nursing room exists, I'm not sure whether I'd want a live patch-in of the conference though, since it could distract the babe."* and *"I love the idea of a nursing mothers room with video hook up. I missed two sessions a day for 4 CHI conferences running back to my room to pump."*

When asked about how much childcare was needed on a daily basis, the top three responses were: over multiple talk sessions (49.4%), a full day (31.7%), and the full day and evening (8.9%). The hours of care is also an important related consideration. For example, many social events important to professional networking occur in the evenings so it was suggested that we may want to consider offering one evening of care.

**Cost.** When asked about the maximum pay rate for a full day (eight hours) of childcare for one child (a closed-form question ranging from $60 USD/day to $140/day), the most common response was $80/day (27.8% of respondents). Over 69.6% preferred $100/day or less; however, 13.9% of respondents were willing to pay $140 or more (Figure 2). When asked about *who* would pay for childcare—for example, the caretaker or the employer—72.2% responded paying themselves "out-of-pocket" and only 12% thought that they could get some or all of the childcare costs covered by their institution (14% were not sure).

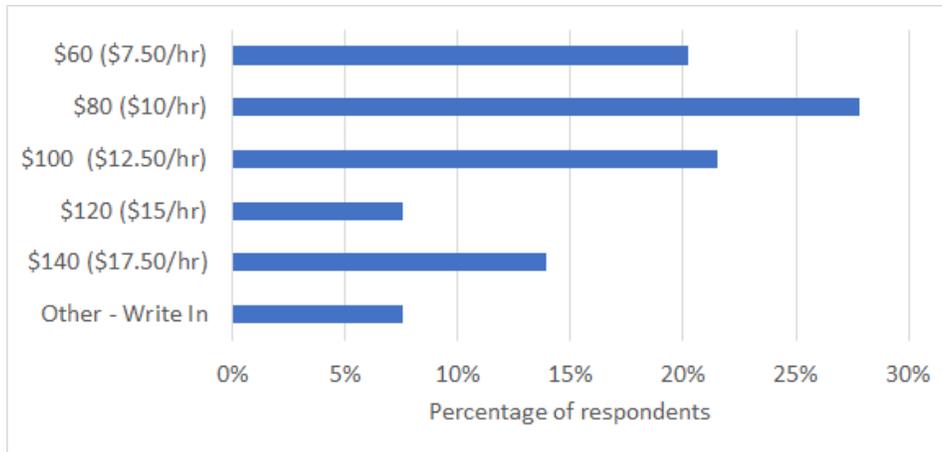

Figure 2. Maximum possible pay rate for 8 hours of childcare (*N*=95)

Clearly, cost is an important factor that disproportionately affects students or others with lower income, those who cannot reimburse childcare, and international attendees with higher travel costs. For example, one respondent said: "*For me, I'm almost always traveling internationally to CHI and because I'm not (a) super rich or (b) part of a couple where both partners are in HCI, it seems pretty unlikely that I'll be able to afford to bring my kids to CHI ever anyway.*"

Others felt that how childcare costs are handled is an important cultural signal to attendees and beyond. For example, a few respondents explicitly stated that SIGCHI or ACM should subsidize these costs, such as "*I would also opt for a free child care and have it paid by SIGCHI as a positive sign of the SIGCHI community and the multimillion conference.*" Another respondent said that s/he "*will ask my department to reimburse the cost (at least for the daycare there) to also signal that these cost do exist and that this is an important issue that not only needs to be taken care of and needs money.*"

Over email, a person suggested that we try and solicit corporate sponsorships for childcare—a model successfully employed at other conferences: "*One time, we paid about $50/day [for childcare]. But all the other times, it's been free. Typically, it is subsidized by corporate sponsors. Microsoft sponsored one I went to, and an academic sponsor did [Grace Hopper Conference]. I think that major companies are likely to be willing to sponsor it.*"

**Language preferences.** As an international conference, we did not want to presume that English would meet every child's need. Indeed, five respondents stated that they would prefer a different spoken language if possible (3 German, 1 Korean, 1 French).

**Childcare Advertisement.** We received this comment by email, which summarizes well the need to advertise early and broadly. "*One thing really important is to advertise very early. The community has started figuring out that this would be really useful to parents, but it often gets organized last minute, and many people have already made their childcare plans. Then it appears there isn't much demand, when in fact there is a huge demand. Also, at Grace Hopper, I met a bunch of people who didn't even know there was childcare. So plan early and make everyone aware of it so that you can get the most bang for your buck. Once you get an outfit and a room and anything, you want to fill that room.*"

**Survey Limitations.** Our survey had multiple limitations. First, there were a low number of students and industry respondents. This could be a reflection of how we advertised our survey and the audience it reached or simply that this population is less likely to bring children to conferences. Second, we did not include a no-cost-to-parent ($0) option when asking about maximum preferred per-day childcare costs. Finally, our survey was primarily aimed at respondents who desired to bring children to CHI2018 in Montreal, Canada—those respondents who indicated otherwise were not given the same questions. While we believe many of our findings have broad applicability, each academic field as well as host city and country present unique challenges that need to be addressed.

## Making Decisions about Family Support Services

We identified two possible services for childcare: a local babysitting service that works with major hotels in Montreal and an American company that organizes childcare service at international conferences, KiddieCorp, which has worked with ACM and SIGCHI in the past. We brought these recommendations with the conference's logistical team to discuss organizational and policy considerations before making a final decision.

**Financial considerations**. The CHI general chairs are given some autonomy in how the budget is allocated, but because it is one of the largest ACM conferences, there is also significant oversight on large budgetary decisions. While the budget is in the multi-million dollar range, by the time the general chairs are brought into the decision-making process, millions of that budget are already allocated to things like venue costs, food and beverage, and general logistics of running the conference, which limits flexibility.

As general chairs, we allocated $15,000 to subsidize on-site childcare, which was approved by the SIGCHI Steering Committee and ACM. Because recent iterations of childcare have had low uptake, we justified our decision from the survey data above. In addition, we argued our shared desire to provide what was considered an important service by all parties. The conference organization recognized that providing such a service would be particularly beneficial to women and women of color [4].

**Space considerations**. With up to 23 parallel tracks, the scale of CHI is such that a significant amount of planning is required to fit all of the sessions into the space available. Finding space for non-program activities, such as on-site childcare and a nursing room, adds to this complexity. We were fortunate to have a contract with sufficient space for childcare throughout the week; however, because the weekend activities (primarily workshops) require less space, we had fewer rooms available, and no appropriate available room for this type of service. We allocated space for a nursing room throughout the conference.

**Liability concerns**. Our parent organization, ACM, is required to pay insurance to cover any liability for anything that occurs on the premises of the conference: the addition of on-site family services introduces some additional liability concerns. In particular, access to the site requires a badge to track who and how many people are present. This meant that every child that entered the conference venue required a pass. Moreover, the liability insurance provided in part by the ACM and in part by the venue itself does not cover potential harm that could come to children, such as a child harming themselves with a toy or burning themselves on a bottle warmer provided by the conference organizers in a nursing room. In our case, the larger liability coverage provided by KiddieCorp was appealing to the ACM and the logistical team, as it eased concerns around potential harm that could come to children interacting in a space that was not designed for them specifically. Finally, organizers also took into account the presence of children at events regarding the local alcohol laws regarding sales and services.

**Organizational concerns.** Finally, when making decisions, we also considered the level of organization required by the conference to organize the childcare process (*e.g.,* schedules, advanced registrations, on-site registrations, drop off/pick ups, toys and activities): the babysitting service would have required the conference to organize this process, while KiddieCorp includes it in their services.

## Running Family Support Services

Based on parents' needs and constraints voiced during the survey, general advice and organizational constraints, the conference and family chairs decided to support attendees with families through multiple options, including a child pass, on-site childcare services, and a nursing room. We hoped this flexibility enabled broad attendance to CHI 2018 by attendees with families.

**Child Pass:** To provide access to the conference center and to make kids feel welcome at CHI 2018, we included a $10 "child pass" for kids 0–18 to accompany their parents. This pass provided access to the convention center and to the reception for children accompanied by their parents. We used the child pass registrations to communicate information to parents.

**Childcare:** We supported on-site childcare provided by KiddieCorp, a professional on-site childcare service. Children had a large common room with toys, snacks, and activities such as crafts, construction toys, tiny tikes toys, books. The KiddieCorp team members are uniformed, qualified, screened, and experienced employees who have completed the KiddieCorp training program. KiddieCorp offers childcare in English, for children 6 months to 12 years old.

Childcare was available during the four days of the main conference (not during the workshops due to space constraints) at the cost of $10/hour USD. Childcare was offered during technical sessions, but not during the lunch break or evening networking events. Parents indicated interest in childcare through their conference registration and registered for childcare on the KiddieCorp site. They also could register kids on-site.

**Nursing room:** We provided a quiet, semi-private nursing room for feeding and changing. The nursing room had comfortable sofas, a changing table, nursing chairs, a play mat, a kettle, and a fridge to store milk. There were signs in the nursing room that directed caregivers where to view live streams of the paper sessions that could be accessed on a laptop or mobile device.

**Other Options:** We also suggested a local babysitting service for attendees looking for babysitting services beyond what the conference could offer (*e.g.,* for evening events) and linked to the HCI parent Facebook group[3] so parents could get in touch with each other.

## Post-mortem on Family Support Services

Family support services at CHI 2018 enabled 61 kids to accompany their parents at the conference with the conference badges. Many attendees commented on the higher proportion of kids around the conference site, typically quite positively.

Specifically, 24 kids from 17 families used the on-site childcare services. The service was full on the first two days of the conference and used at over 80% capacity for the last two. While we had originally planned for 22 kids and 5 providers, we reduced the number of providers at the deadline stated on the KiddieCorp contract, based on the registrations from the advanced registration period that ended a month before the conference. In the end, we contracted the daycare services to 18 concurrent kids with 4 daycare providers for each block of time.

---

[3] https://www.facebook.com/groups/779317265507577/

To gain an understanding of how childcare services at CHI were used and perceived, we gathered data from two post-conference surveys: first, childcare-related questions were included in the official post-CHI survey sent to the nearly 3,400 attendees; second, we conducted our own survey, which was advertised similarly to our pre-conference survey and also emailed directly to attendees who registered for a child pass and/or indicated interest for childcare at the conference. Both surveys were launched just after the conference and remained open for several weeks. All questions were optional, so percentages indicate proportion of respondents for each question.

## Official Post-CHI Survey Findings

Of the 3,372 registered attendees, 692 filled out the survey. For demographics (*N*=604), 54.5% were male, 40.7% female, 1.2% transgender, gender variant or non-conforming, and 3.6% selected "preferred not to answer." The most common profession was "student" selected by 47.6% (285 of 598) of respondents followed by professor (21.4%), researcher (18.9%), and other (12.0%). For work setting, 81.5% (488 of 599) selected university followed by industry research lab (9.5%), practitioner (5.5%), government-funded institution (2.8%), and consultant (0.67%).

A majority of respondents felt that it was important that CHI offer childcare-related services (whether the respondent personally used them or not): 63.0% (384 of 610) selected *very* or *extremely* important for "nursing room" followed by 61.1% (376/615) for "on-site childcare" and 46.5% (282/606) for "child passes". Similarly, a majority felt that the CHI conference is "family-friendly": 63.1% (364/577) agreed and 8.0% disagreed (46/577). Finally, when asked about what services the respondent would expect to use at CHI 2019, 20% (43 of 215) desired on-site childcare and 10.7% wanted a nursing or pumping room.

## Family Chairs Post-Conference Survey Findings

In total, we had 62 respondents, including 33 female, 24 male, and two nonbinary or gender noncomforming; 47 reported being parents (76%). 22 respondents stated bringing their children to the conference. Of these, over three quarters stated that the family support services made a difference in being able to attend the conference (17). Of parents that did not bring children, half reported considering it before ultimately deciding no. When asked about this decision process (closed-form), the top three reasons selected for bringing a child to CHI2018 included: not having a good caregiving/childcare option at home (32%), enjoying travel with children (18%), and providing an enriching educational experience (18%). The top three reasons for not bringing children, included: travel costs (45%), preferring not to travel professionally with children (32%), or that the services were considered inappropriate for the child's age (32%). See Table 1.

Table 1. Reasons for bringing, or not bringing their kids to the conference

| **Top reasons for bringing kids (*N*=22)** (single-selection response) | **Top reasons for not bringing kids (*N*=22)** (multi-selection response) |
|---|---|
| ● I do not have good caregiving/childcare options at home (32%)<br>● I enjoy traveling with my children (18%)<br>● I thought it would be an educational experience (18%).<br>● We made a family vacation out of the trip (14%) | ● Travel costs (too expensive) (45%)<br>● I don't like to travel professionally with my children (32%)<br>● I felt that my children were too old (32%)<br>● Childcare costs (too expensive) (18%)<br>● Location of conference (14%)<br>● I felt that my children were too young (14%) |
| ● I am currently nursing (9%)<br>● I am a single parent (5%)<br>● I want to maximize my time with my children (5%) | ● I cannot take my kids out of school during the term (18%)<br>● My kids are not native English speakers (14%) |

**Services used and reactions.** When asked if they felt that children were welcome at the CHI2018 conference: 79% (49/62) agreed, 11% (7) did "not know", and 6.5% (4) selected "not really." When asked about services used at CHI 2018, 27% (17 of 62) said the child pass, 23% (14) the on-site childcare, 6% (4) the nursing room. In general, parents were satisfied with the services provided: all parents (14 of 14) were *satisfied* or *very satisfied* with the onsite childcare, 83% (14 of 17) with the child pass, and 75% (3 of 4) with the nursing room. For onsite childcare, 53% (7 of 13) desired longer hours. When asked about CHI 2019 in Glasgow, UK, 53.7% (29 of 54) expected to use family support services (if they were available), 31.5% were unlikely, and 14.8% were unsure (Figure 3).

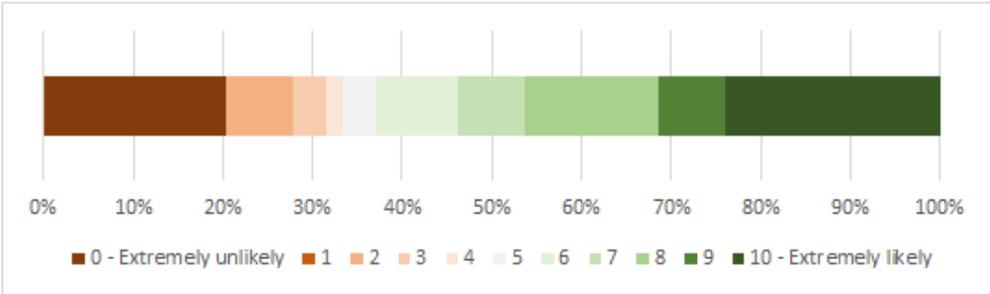

**Figure 3. Distributions of participant answers regarding their use of family support services at next year's conference**

**Open-ended feedback.** We provided space for respondents to leave comments, suggestions, and other feedback about family services at CHI; 41.9% (26) responded. A large majority were positive, thanking us for our efforts and emphasizing the impact these services had on their conference experience: *"Extremely happy with the on-site childcare program this year! My kid also enjoyed her time with the teachers and other kids. This service made a HUGE difference to our travel decision and experiences this time."*, and *"Thank you because of this type of effort, I plan to attend CHI in the future!!!"* Most of the negative comments were about the high cost; one was about safety concerns.

For suggestions, four common themes emerged: reducing cost, publishing more information *a priori* to help with planning (*e.g., "show pictures of the space so parents know what to expect"* or provide a *"cheat sheet on different child-related travel questions such as does the city/country require a child seat for taxi"*), extending childcare hours to cover lunch and pre-conference workshops, and more activities for older children (5+) similar to CHIkids in the late 1990s (*e.g.,* a CHI session for children to prepare and present a talk on an HCI topic, co-design sessions with children, marking specific demos as "kid friendly" on the event map).

# Reflections and Recommendations

> *"I'm not a parent, nor do I intend to be, but [childcare] is a critical service to our community, especially for students, junior scholars, single parents, and others who may otherwise not be able to attend (and if they can't attend, they can't publish in the proceedings)."*

There are several issues we encountered when attempting to integrate family support services into the CHI 2018 conference, some of which may be specific to our community and culture, and some that likely will apply more generally. We reflect on the process and provide recommendations to encourage and help facilitate other conferences to integrate family support services.

**Get support.** To provide family services, it is critical to receive ample support, of different types (financial, logistical, even philosophical), and from different sources. Financial support includes discussing daycare costs, backing, sponsorship, insurance and liability issues whereas logistical support includes booking rooms and organizing registration. Having general conference chairs supportive of providing family-friendly options is critical to help implement options to make it easier for attendees with families. We also recommend getting support from the community: in our case, they answered our survey enthusiastically, which gave us the tools to organize childcare.

**Have Family Support Chairs.** In addition to making it easier from an organizational perspective, having dedicated co-chairs announce clearly that the conference is dedicated to supporting its attendees with caregiving needs. Family support chairs can assess the needs of the community and possibilities, source options, and implement the vision for and logistics of providing support for families. They can also ensure that the services are running smoothly during the conference, checking in with the childcare staff a few times a day, and talking to parents to address concerns.

**Give it time.** Organizing family services is an ongoing process. It would probably take 3 years of continuous service offering before we can truly know the use of the services, as it might depend on the location of the conference, and that people need to know in advance whether to bring children at conferences.

**Advertise early and broadly** about care services so that people have a chance to plan travel. Some attendees might even take childcare into account when submitting papers. Information should be available on your website, with regular social media reminders. It is important to try to reach diverse groups. Look for social media groups for parents in your research community.

**Welcome children to events.** We recommend welcoming children to social events, demonstrations and exhibit halls. Explicitly check in advance with offsite venues about their child policies to ensure that kids are welcomed at official conference events. If they are not, we recommend announcing in advance when events can only be attended by adults, so parents can plan accordingly.

**Think about students.** While most of the attendees that took advantage of childcare were faculty members, it is important to think about accommodating students better, whether with subsidized childcare cost, or organizing the logistics of having a parent also do student volunteering duties.

**Support kids of all ages.** Parents with kids of all ages may need support, and the support type varies with age ranges. Consider the following groups when planning your services: babies (0–1), toddlers (1–5) and older kids (6+).

**Provide a nursing room**: Parents need a quiet, semi-private room to feed their kids. It can also be used for babies napping, or a place for them to calm down. We recommend placing a mini-fridge in this room to support caretakers to store milk, a bottle warmer or a kettle, a changing pad, and a closed garbage can. We recommend having a video feed to the conference sessions.

**Provide subsidized childcare services**. While not accommodating all kids, English-language care is sufficient for most. Require that the selected childcare service has good check-in and check-out procedures. The **hours of childcare** should at minimum cover the technical sessions, plus 15 minutes before and after the sessions so parents can listen to the entire session. Additional hours to cover lunch and evening events would likely be appreciated and used by some parents. Parents should have the ability to sign up for individual hours, or for blocks of time (half-days). **Cost** should be subsidized with a sponsor if possible. Otherwise, while we found a large variation in what people were willing to pay, $10/hr USD per child seems most accepted. Advanced **Sign up** is recommended so the conference can plan and adjust to the demand. To help with advance registrations, consider an early-bird childcare registration discount. We also recommend a partial deposit during the advanced registration, which would ensure that services booked are used, yet that the balance paid on-site considers the actual use. The option for on-site sign ups is necessary.

**Offer registration discounts** for kids and for caretakers. A child **badge** typically lets them be covered under the conference insurance policy. It can also give them access to the conference rooms so they can reach their parents and see demos. It also provides them a sense of belonging to the conference. **Caretakers** (spouse, family members, babysitters that are not part of the HCI community) can also benefit from a discounted registration, so they can come on-site with the kids. A complete on-site access is critical, and it should include attendance to evening events (conference reception, demos, etc.). However, access to the rooms with talks may not be necessary.

**Consider offering childcare support at the program committee (PC) meeting.** The PC meetings are another event that attendees with family must navigate. A respondent of our pre-conference survey indicated that "*For me, the PC meeting is far more of a problem as a parent than CHI itself… I've declined PC committee invites for [multiple years] now. Several other parents amongst my colleagues are in the same situation.*" Consider supporting childcare for your meetings to be inclusive of all members of your community.

**Donations.** Conferences can act responsibly by donating all the family support services artifacts bought for the conference to a local women's shelter.

## Towards the future

The ACM SIGCHI community aims to be inclusive and diverse. SIGCHI conferences are taking steps to increase the participation and success of underrepresented groups in HCI. The advice page on organizing a SIGCHI sponsored conference includes a paragraph on the need to develop a policy regarding children at conferences[4]. We encourage other special interest groups to adopt similar policy and encourage their members to consider how to improve access to conferences or include supporting attendees with families.

This conference attendee summarized clearly the need for these services:

---

[4] https://sigchi.org/conferences/organizer-resources/organising-a-sigchi-sponsored-or-co-sponsored-conference/

*"THIS. It is so needed. I am terrified this will be a "one off" (even having a nursing room has varied from year to year). Programs like these make parenting in the community visible and send the right message about participation from primary caregivers who also happen to be HCI researchers. Oh, and I've also made professional connections I otherwise wouldn't have, because we brought our kids and wanted to connect as CHI parents to swap tips AND talk about research!"*

Pictures to accompany the article

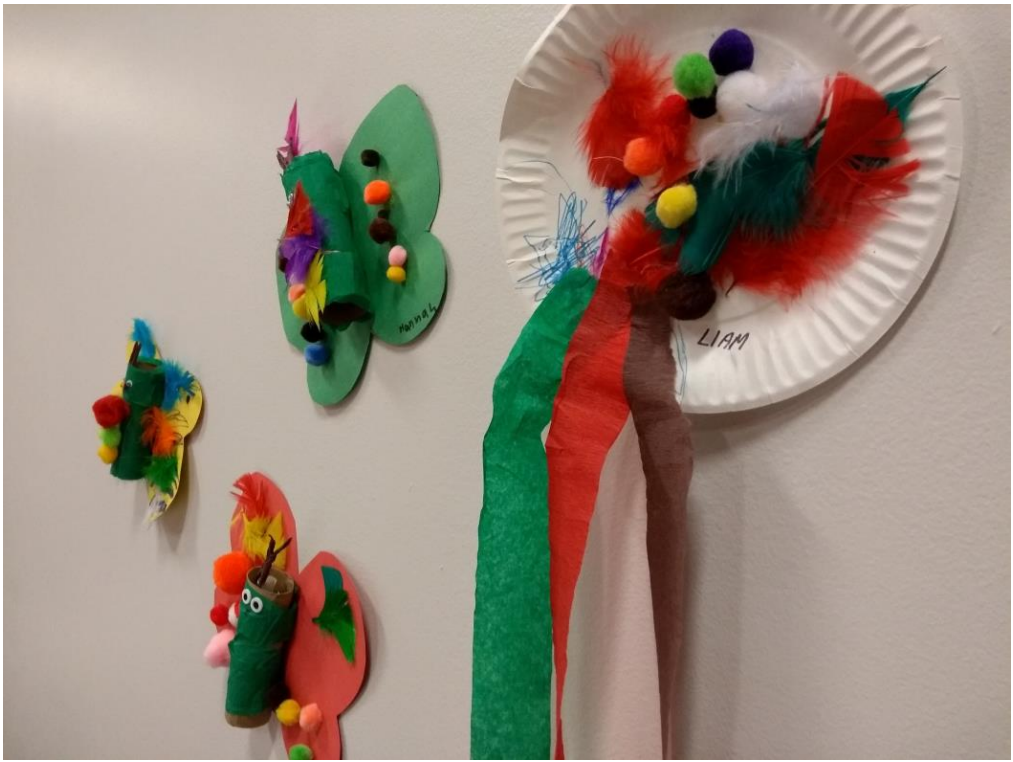

Artwork created by kids while in the CHI childcare

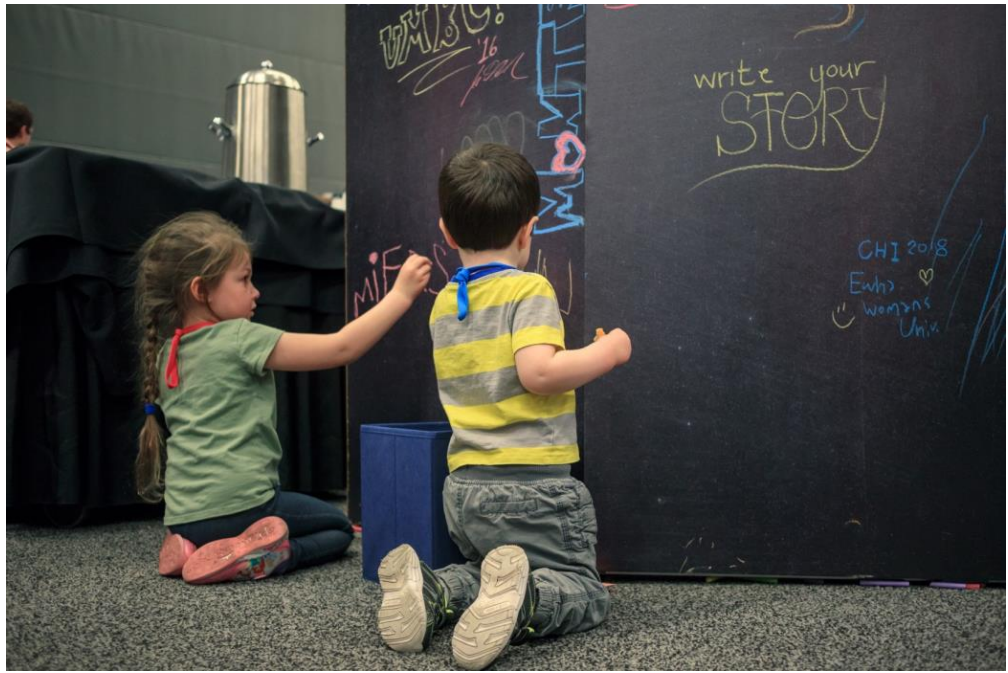

The chalkboard column was quite popular with kids and adults in the exhibit hall.

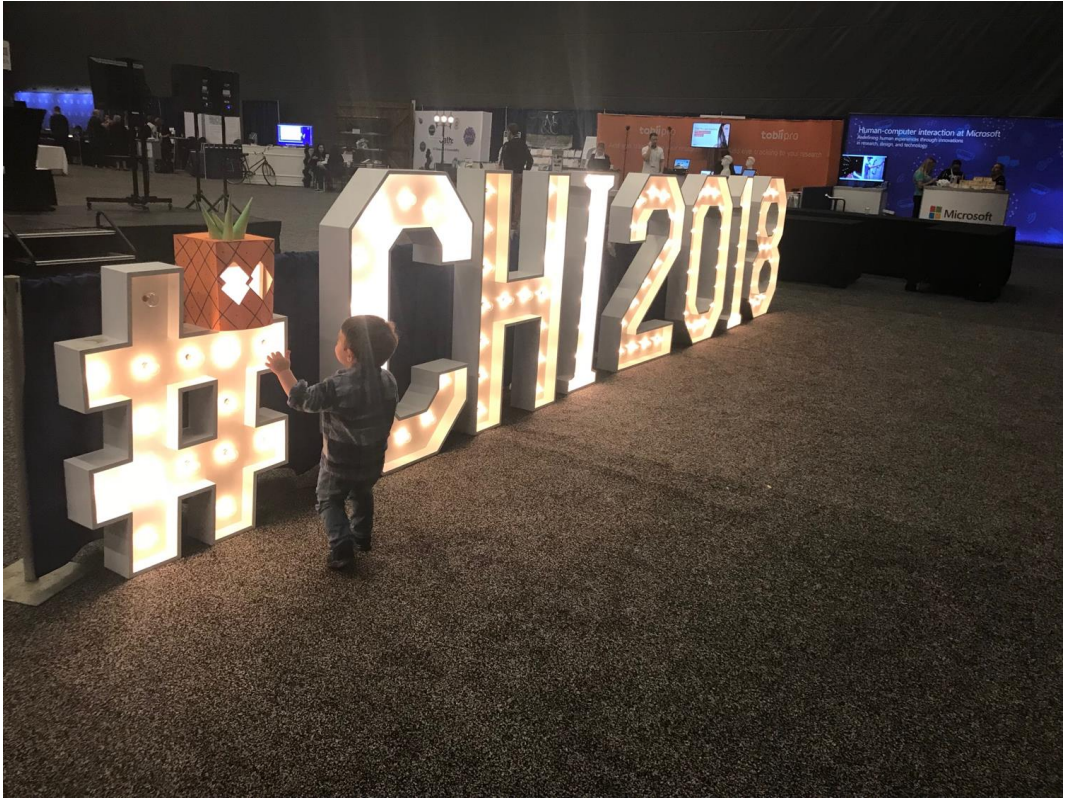

Smaller attendees appreciated the colourful CHI2018 giant hashtag

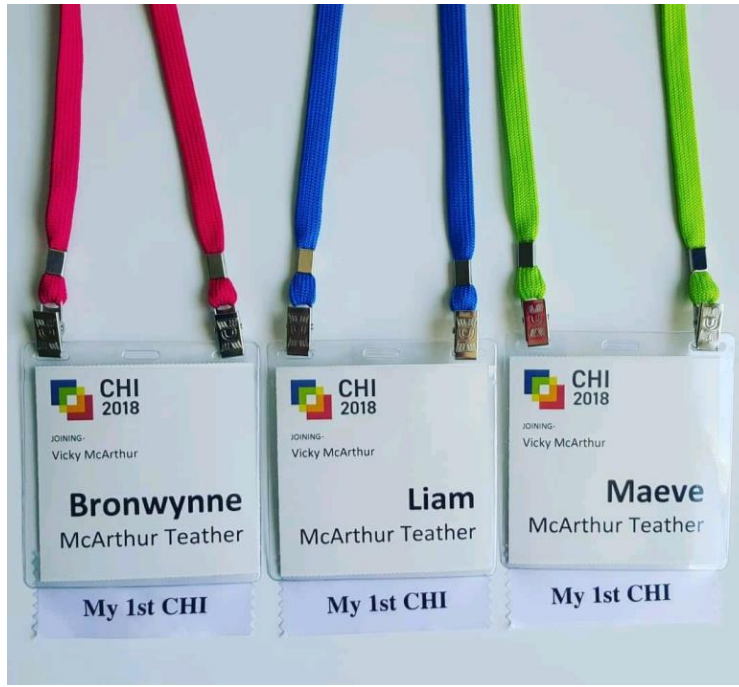

"A lovely idea to let children be part of CHI—I especially liked to see little ones with badges staying 1st CHI!"

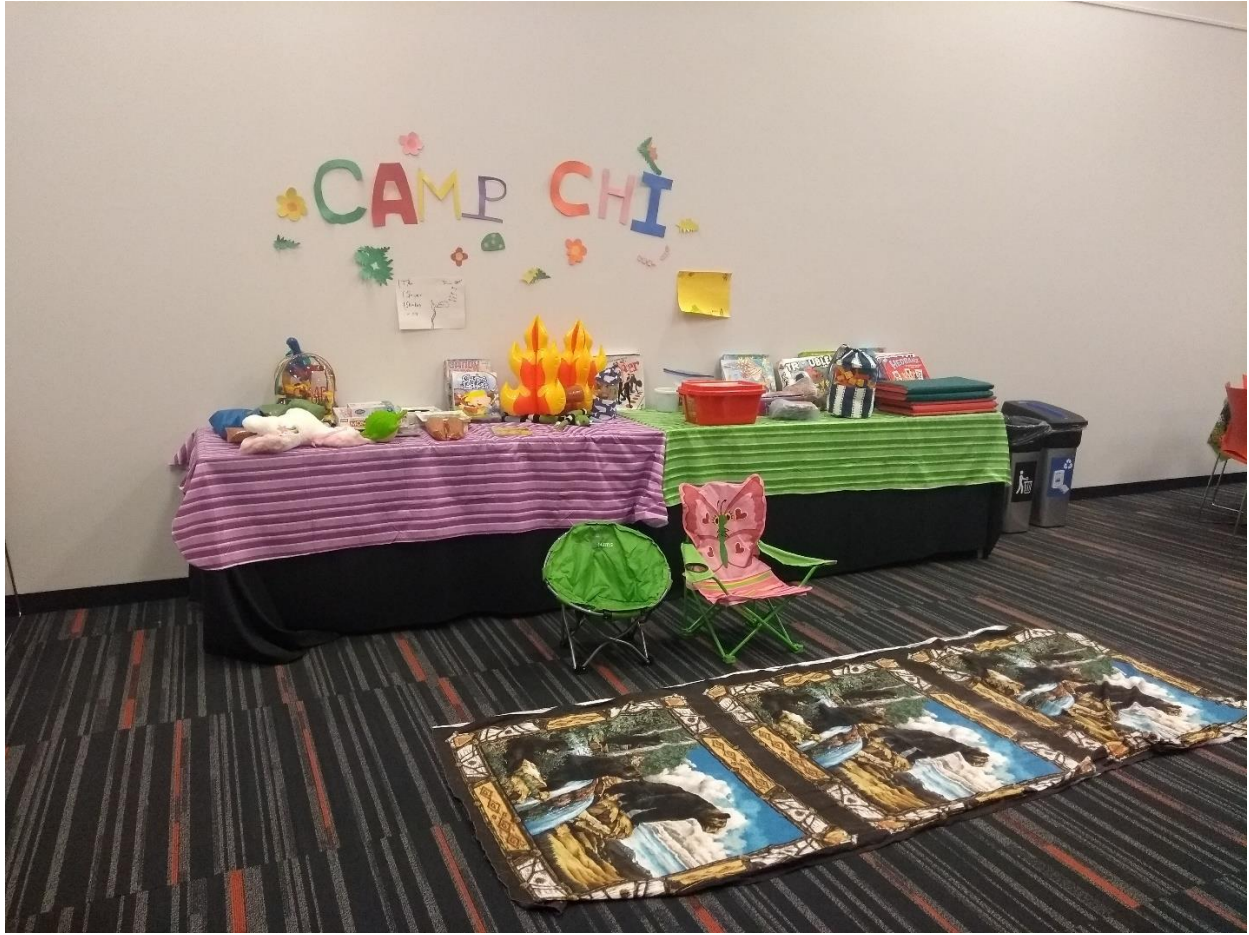

The CHI2018 children's theme was 'camp CHI'